\documentclass[prl,twocolumn,amssymb,showpacs]{revtex4}
 \usepackage{epsfig,amssymb,amsmath}
\begin{document}
 \title{The onset of jamming as the sudden emergence of an infinite
 $k$-core cluster}
 \author{J. M. Schwarz and Andrea J. Liu}
 \affiliation{Department of Chemistry and Biochemistry, UCLA, Los
 Angeles, CA 90095 and Department of Physics and Astronomy, University
 of Pennsylvania, Philadelphia, PA 19104 }
\author{L. Q. Chayes}
 \affiliation{Department of Mathematics, UCLA, Los Angeles, California
 90095
 }
% \pacs{64.60.Ak,64.70.Pf,83.80.Fg}
%\begin{document}
%\maketitle
 \begin{abstract}
 A theory is constructed to describe the zero-temperature jamming transition 
of repulsive soft spheres as the
 density is increased.
 Local mechanical stability imposes a constraint on the minimum number
 of bonds per
 particle; we argue that this constraint suggests an analogy to $k$-core
 percolation.
 The latter model can be
 solved exactly on the Bethe lattice, and the resulting transition has a
 mixed
 first-order/continuous character reminiscent of the jamming transition.  In particular, the exponents
 characterizing the continuous parts of both transitions appear to be the same.
Finally, numerical simulations suggest that in
 finite
 dimensions the $k$-core transition can be discontinuous with a
 nontrivial
 diverging
 correlation length.
 \end{abstract}
\pacs{64.60.Ak, 64.70.Pf, 83.80.Fg}
\maketitle
%\pacs{64.60.Ak,64.70.Pf,83.80.Fg}
 Understanding a continuous phase transition is tantamount to
 determining the universality class
 to which it belongs.  In contrast, understanding the nature of a
 discontinuous change of phase requires a
 detailed study of the system at hand.
 Under normal
 circumstances~\cite{Liggett}, the two categories are mutually exclusive.
 However, there are a few examples of continuous transitions
 that exhibit characteristics of first-order
 transitions~\cite
 {Thouless,ACCN,GKS,kirkpatrick.wolynes,kirkpatrick.thirumalai,
 fredrickson.andersen,kob.andersen,toninelli.biroli.fisher,sellitto.biroli.toninelli}.
 In this Letter, we will present arguments that
 the jamming transition
 in sphere
 packings\cite
 {ohern.langer.liu.nagel,ohern.silbert.liu.nagel,silbert.liu.nagel}
belongs to this class and
 can genuinely be described as
 both continuous and discontinuous.  Indeed, we will identify the
 minimal physics needed to capture the nature of the transition by
 analogy to the
 $k$-core percolation model, and show by exact calculation that the
 latter model has a true mixed transition of this type with similar exponents
 at the level of mean-field theory.  We also present numerical evidence that $k$-core
models can still exhibit mixed transitions in finite dimensions.  We remark that, 
starting from a different vantage point, Toninelli, {\it et al.}~\cite{recent} have arrived at a model
of the $k$-core type and have reached similar conclusions about the nature of the transition
in their studies of kinetically-constrained models.

 Numerical
 studies~\cite
 {ohern.langer.liu.nagel,ohern.silbert.liu.nagel,silbert.liu.nagel}
 of sphere packings at zero
 temperature suggest that there is a packing density $\phi_{c}$
 (Point J) where
 the onset of jamming is truly sharp; {i.e.} the static bulk
 and shear moduli vanish for $\phi\le\phi_c$ and are nonzero for
 $\phi>\phi_c$.  This transition exists for spheres that repel
 when they overlap and otherwise do not interact.
 For small $\phi$, particles easily arrange themselves so as not to overlap with any
 other particle and hence the total potential energy is $V \equiv 0$.
 As $\phi$ is increased, there is a particular value of $\phi_c$
 above which the
 particles can no longer ``avoid'' each other and $V$ becomes nonzero.
 The average coordination number (the average number of
 overlapping neighbors per particle) is $Z=0$ for $\phi<\phi_{c}$. As
 $\phi$ approaches
 $\phi_{c}$ from above, however, the behavior is very different:
 $<Z> \approx Z_{c}+Z_{0} (\phi-\phi_c)^\beta$, where
 $\beta=0.49 \pm 0.04$~\cite{ohern.silbert.liu.nagel}.  Moreover, the
 singular part of
 the shear modulus
 vanishes with the
 exponent $\gamma=0.48 \pm 0.05$~\cite{ohern.silbert.liu.nagel}
  and recent
 simulations by
  Silbert, {\it et al.}~\cite{silbert.liu.nagel} find a diverging length scale
exponent
 $\nu=0.24 \pm 0.03$ \cite{silbert.liu.nagel,reichhardt}.

 These numerical results imply that the transition at Point  J
 has characteristics of both
 types: certainly there is a discontinuity in the
 average coordination number, $\langle Z \rangle$, but
 as
 the transition is approached from the ordered
 (jammed) phase, it
 exhibits the typical singularities associated with
 continuous transitions; $\langle Z \rangle$ tends to its limiting
 value with a nontrivial power-law and there are divergent length scales.

 We will now present arguments that the Point J transition is indeed a
 mixed transition, many aspects of which can properly be understood by
 analogy to a
 relatively simple
 model called ``$k$-core percolation"  (sometimes also called ``bootstrap
 percolation" \cite{chalupa.leath.reich}).
 Let us start with an informal discussion of the essentials of the
  jamming model.
 Clearly, a jammed packing of spheres at $T=0$ must be mechanically
 stable.
 For a sphere in $d$ dimensions to be locally stable, it must have
 interactions
 ({\it i.~e.}\ overlap)
 with at least $d+1$ neighboring spheres~\cite{shlomo}.
 Evidently, spheres with fewer than $d+1$ overlapping neighbors do not
 contribute to the formation of a jammed
 structure and thus are {\it irrelevant}.  Thus we may envision the
 mechanics for a system
 below the jamming threshold density as its energy
 is lowered towards the minimum:  although large
 clusters of overlapping particles may happen to form, those at the
 boundary of the cluster
 are unstable and will move away, further lowering the energy.  This in
 turn exposes secondary particles, who are in turn forced to move away,
 and so forth
 until the cluster
 dissolves.   At high density the situation is more complicated.
 However, it is still
 true that all particles
 that {\it do} contribute to
 the jammed structure must have at least $d+1$ overlapping
 neighbors that are not ``irrelevant'', and each of these overlapping
 neighbors must have
 at least $d+1$ overlapping neighbors that are not irrelevant, and so
 on.  In other words, only particles that survive this entire hierarchy
 of irrelevance can
 contribute to the jammed structure.

 These considerations are suggestive of the $k$-core percolation model,
 defined as follows.
 Consider
 a regular lattice of coordination number $Z_{\rm max}$ and some
 integer $k$ with $2\le k< Z_{\rm max}$.  Initially, sites are
 independently occupied
 with probability $p$.  In the first
 stage, all occupied sites
 with fewer than $k$ neighboring occupied sites are eliminated.  Then,
 this
 decimation process is
 applied to the surviving occupied sites, and so on, until all surviving
 sites (if any)
 have at least $k$ surviving neighbors.  Thus, at the end of this
 process, every
 surviving site has at least $k$
 neighbors, all of whom in turn have at
 least $k$ neighbors, etc. The surviving sites are
 called the $k$-core and  phases of the model are determined
 by the presence or absence of an infinite cluster of these survivors.

 The overall analogy between the two models is self-evident.
 The initiating density $p$ corresponds the the packing fraction $\phi$,
 $k$ corresponds to $d+1$ and $Z_{\rm max}$ to the so-called
        {\it kissing number}, that is
 the maximum number of equivalent hyperspheres in $d$ dimensions
 that can touch a central one without overlaps.
 (In $d=2$, $Z_{\rm max}=6$, and in $d=3$, $Z_{\rm max}=12$.)

 In the mean-field theory of $k$-core percolation, i.e.
 the Bethe lattice and infinite-range complete graph models, it is
 well-established that the order parameter undergoes a discontinuous
 jump at threshold \cite{chalupa.leath.reich,pittel.spencer.wormald},
 accompanied
 by a square-root
 singularity~\cite
 {chalupa.leath.reich,pittel.spencer.wormald,moukarzel.duxbury.leath}.
 However, the fact that the latter was indicative of
 a {\em critical phenomenon} has heretofore been underemphasized; in particular,
 a divergent length scale had not been identified.
 Below we will show that, at least on the Bethe lattice, there is indeed
 critical behavior in the sense of a diverging susceptibility and
correlation length and the various exponents
 are in (rather dramatic) agreement with their counterparts in the Point
 J simulations.

 The $k$-core percolation model can be solved exactly on the Bethe
 lattice.
 We begin by considering the half-space Bethe lattice, for which
 we derive recursion relations for quantities at level
 $n+1$
 in terms of quantities at level $n$.  All occupied sites at level 0 of
 the half-space Bethe lattice belong to what we call the ``deep core."  We keep track
 of two quantities,
 the probability of belonging to the deep core at level $n+1$,
 $\Upsilon^{\rm HS}_{n+1}$,
 which requires the site at level $n+1$ to have at least $k$ neighbors
 at level $n$
 that belong to either the deep core or to what we call the ``corona."
 To be in the corona at level $n+1$, a site must have {\it exactly}
 $k-1$ neighbors at
 level $n$ that belong to the deep core or the corona.
 We denote by $\Phi^{\rm HS}_{n+1}$ the probability of belonging to the 
 corona at level $n+1$,
 and by $\Gamma^{\rm HS}_{n+1} \equiv \Upsilon^{\rm HS}_{n+1}+\Phi^{\rm HS}_{n+1}$
 the probability
 of belonging to either the deep core or the
 corona.
 The deep core will necessarily be part of the $k$-core; we need to keep
 track of the corona
 because when two half-spaces are glued together to form the full Bethe
 lattice, corona
 can be converted to $k$-core.
 The recursion relation is
 \begin{eqnarray}
 \Gamma^{\rm HS}_{n+1}&=&p\sum_{l=k-1}^{Z_{\rm max}-1} \binom{Z_{\rm
 max}-1}{l}(\Gamma^{\rm HS}_n)^l (1-\Gamma^{\rm HS}_n)^{Z{\rm max}-1-l}
 \nonumber\\
 &\equiv&p\Pi_{k-1}^{Z_{\rm max}}(\Gamma^{\rm HS}_n). \label{recursion}
 \end{eqnarray}
 In the limit of large $n$,
 $\Gamma^{\rm HS}_n=\Gamma^{\rm HS}_{n+1} \equiv \Gamma^{\rm HS}$.
 Clearly, $\Gamma^{\rm HS}=0$ is always a solution.
 However, there can be a nontrivial solution for $p$ exceeding some
 $p_c$.
 For $k \ge 3$ ~\cite{chalupa.leath.reich,moukarzel.duxbury.leath}
 \begin{equation}
 \Gamma^{\rm HS} \sim a + b(p-p_c)^{1/2}
 \label{orderparam}
 \end{equation}
At the transition,
 the curve $p\Pi_{k-1}(\Gamma^{\rm HS})$ is just tangent to 
$\Gamma^{\rm HS}$, {\it i.~e.}\ $p_c\Pi_{k-1}'(\Gamma^{\rm HS})=1$.

 The average coordination number, susceptibility and correlation
 length exponents, which are needed for the comparison to sphere
 packings, must be calculated on the full Bethe
 lattice, obtained by connecting two half-space lattices.
 The
 resulting probability of belonging to the $k$-core, $K$, is
 given by $K=\Upsilon^{\rm HS} +\Phi^{\rm HS}\Gamma^{\rm HS}$.
 This has the same behavior as $\Gamma^{\rm HS}$ in Eq. 2.  The average
 number of
occupied 
 neighboring sites per occupied site ({\it i.~.e.} the average
 coordination number) also behaves in the same fashion as $\Gamma^{\rm
 HS}$ (Eq.~\ref{orderparam}).  It jumps from zero
 for $p<p_c$ to $\langle Z \rangle \approx Z_c + Z_0 (p-p_c)^{1/2}$ for
 $p > p_c$, in excellent agreement with the numerical results for
 sphere packings~\cite{ohern.silbert.liu.nagel}.

 The susceptibility is the sum of correlation functions, $\tau_{\ell,m}
$, connecting levels $\ell$ and $m$ of the Bethe lattice, and has the
form $\chi=\sum_n (Z_{\rm max}-1)^n \tau_{0,n}$.  We consider two
different correlation functions: $\tau_{0,n}^{\#}$ represents the
probability that both level $0$ and level $n$ are connected to the
deep core, while $\tau_{0,n}^{*}$ represents the probability that
levels $0$ and $n$ are connected to each other via the corona~\cite
{chayes.liu.schwarz}.  This latter probability can be derived by
considering the chain of sites connecting a given site at level $0$
to a site at level $n$.  To belong to the corona, each site along the
chain must be occupied and have exactly $k$ neighbors (including the
two adjacent sites along the chain) connected to the deep core or
corona.  The corresponding probability for each site is $\Theta=p
\binom {Z_{\rm max}-2}{k-2}(\Gamma^{\rm HS})^{k-2}(1-\Gamma^{\rm HS})^
{Z_{\rm max}-k}$.  The final probability $\tau_{0,n}^{*}$ therefore
scales as $\Theta^n$.  When this is summed over $n$, it yields a
susceptibility exponent of $\gamma^*=1/2$.  By somewhat more complicated 
reasoning, the
dominant contribution to $\tau_{0,n}^{\#}$ scales as $n \Theta^n$;
this leads to $\gamma^\#=1$.  Note that $\tau_{0,n}^{*}$ measures the
size of the corona, which is the region that can be converted into $k
$-core or not, depending on the state of only one site.  This is the source of cooperativity underlying the
transition.

 Another way to compute a susceptibility is to calculate the response to
 a perturbation;
 in the case of percolation, this corresponds to the addition of 
low-probability ``short routes to the infinite cluster.''  In our system we
 have done this in two ways: first by providing
 a small fraction of additional random
 sites with $k-2$ occupied neighbors that are connected directly to the
 deep core plus corona, and secondly by declaring a small
 fraction of sites to be in the deep core plus corona regardless of their
 connectivity.  Both prescriptions
 yield $\gamma^*=1/2$, as well
 as a ``magnetic field" exponent of $\delta^*=2$.

 The correlation length corresponding to {\em both} susceptibility
exponents diverges with
 the exponent $\nu^*=1/4$ (although there is a logarthmic difference
 between the two).  We use the usual embedding of a Bethe
 lattice in Euclidean space~\cite{Grimmett} to arrive at this result.
 One would expect the usual mean field relation
$\nu=\gamma/2$ to hold; the exponent $\gamma^\#=1$ may be an artifact of the Bethe lattice.
 However, we also obtain $\nu^{\#}=1/2$ by looking at how quickly the
 order
 parameter approaches its bulk value as a function of distance from the
 boundary.
 
The
exponents $\beta=1/2$, $\gamma^*=1/2$, and $\nu^*=1/4$ are in 
excellent agreement with numerical simulations of
 particle packings near Point J.  However, these simulations are carried
 out in $2$ and $3$ dimensions while the $k$-core 
 calculations correspond to infinite dimensions (the mean-field
 limit). This raises the question of whether the mixed nature of the
$k$-core
 transition can survive in finite dimensions.
 Most studies have focused on particularly
simple systems such as the 2$d$ square and triangular 
 lattices~\cite{brazilians}, some
 3$d$ cubic lattices~\cite{adler} and hypercubic lattices~\cite{toninelli.biroli.fisher}.
 For these simple systems, the transition
 falls into one of two categories:  Either the transition is continuous
 or
 it does not occur until $p = 1$.
 Systems that exhibit continuous transitions
 all contain self-sustaining clusters, {\it i.~e.~}
 clusters that are finite and yet survive the
 decimation process.   For example, for $k=3$ on the 2$d$ triangular
 lattice, the smallest self-sustaining cluster is a fully-occupied 
 hexagon and the $k$-core transition appears to
 correspond to ordinary percolation
 of these hexagons \cite{brazilians}.  Systems that fail to
 exhibit a transition below
 $p=1$ apparently contain
 ``unstable voids" \cite{straley,aizenmann.lebowitz,toninelli.biroli.fisher}
 that lead to
 decimation of
 the entire population whenever $p<1$. 
 \begin{figure}[h]
 \begin{center}
 \epsfxsize=7.05cm
 \epsfysize=5.5cm
 \epsfbox{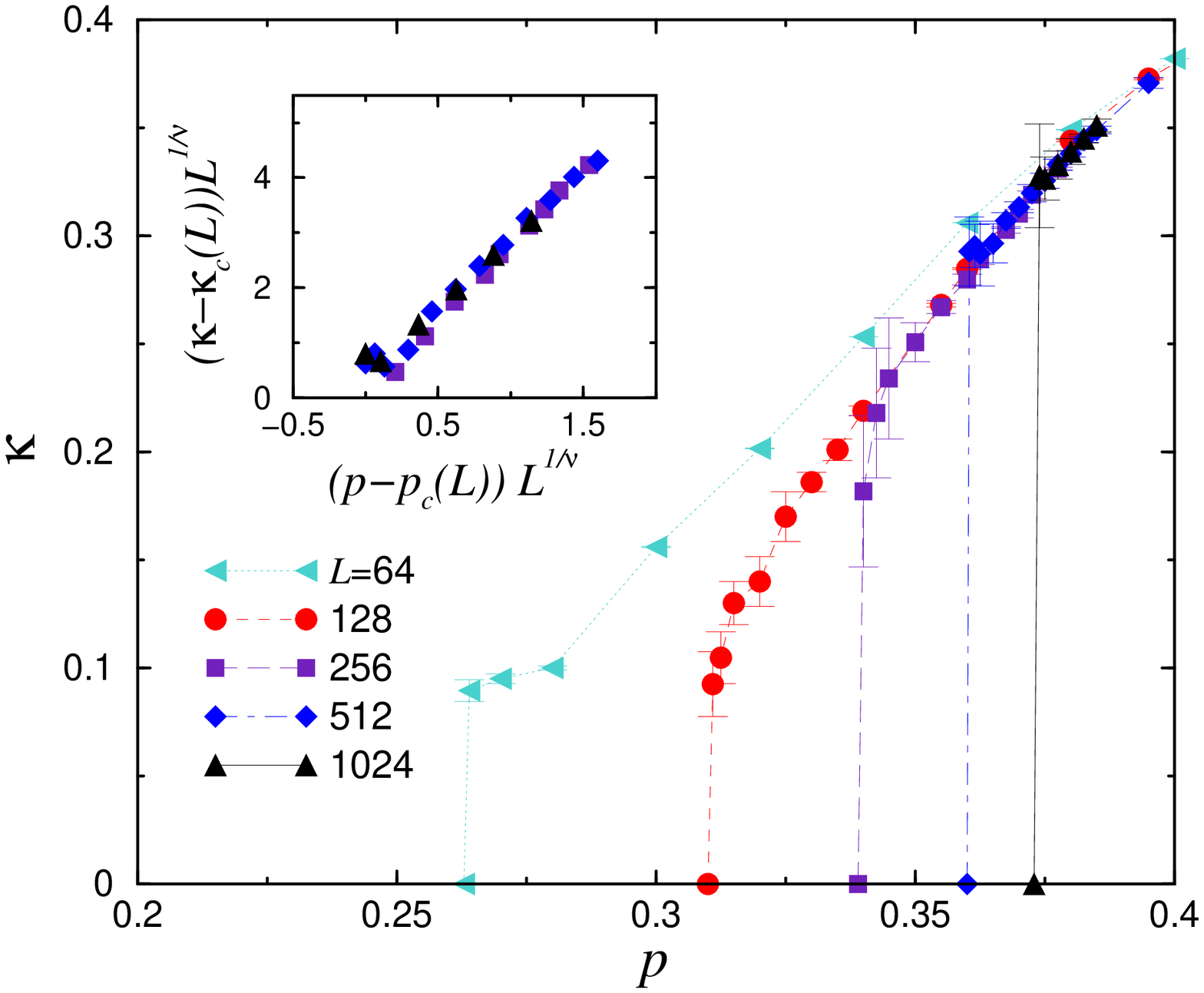}
\hspace{0.2cm}
 \epsfxsize=6.75cm
 \epsfysize=5.4cm
 \epsfbox{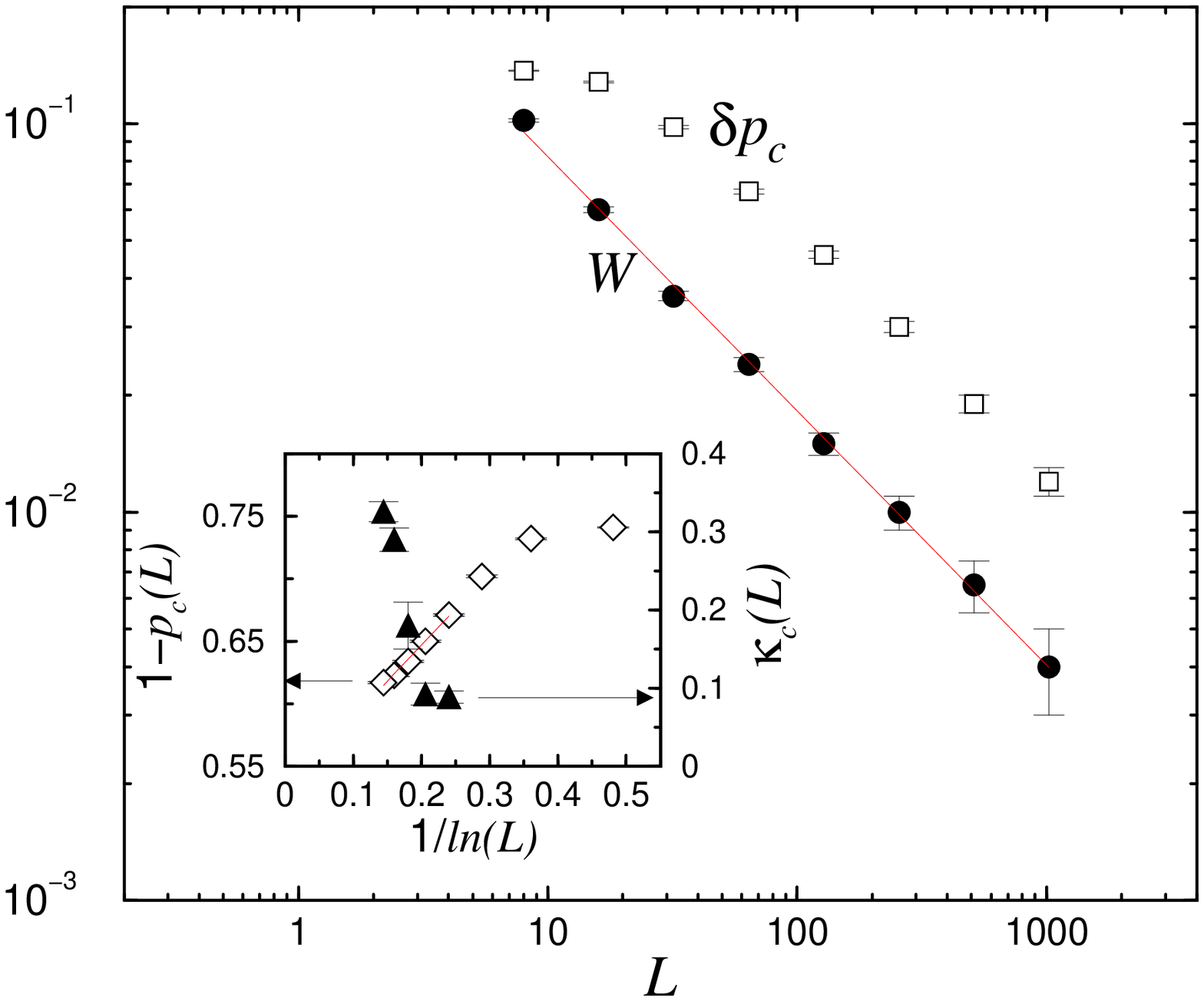}
 \caption{Top: The order parameter $\kappa(p)$ for
 different system lengths $L$.  Inset: Scaling collapse for
 $L=256$ and larger.
 Bottom: Log-log plot of the width of the transition, $W$ (solid circles) and
 the critical point shift $\delta p_c$ (open squares)
 as a function of $L$ with a power-law fit (line)
 with $\nu=1.52(2)$ for $W$ and $\nu=1.53(1)$ for $\delta p_c$. The
 inset shows the jump $\kappa_c$, remains nonzero
 (solid triangles and right axis) and that $p_c$ remains smaller than unity (open diamonds and left axis) as $L \rightarrow \infty$.}
 \end{center}
 \end{figure}

 We regard percolation of self-sustaining clusters
 and unstable voids as ``artifacts'' of simple $k$-core models.
 Indeed, for jammed sphere packings these two effects cannot arise.
 First, self-sustaining clusters of overlapping particles are
 forbidden
 due to the
 repulsive nature of the interactions between particles. Likewise,
 voids (i.e. collections of floaters, or particles with no overlapping neighbors)
 cannot grow because of the force constraints and because floaters can shift around, but cannot actually disappear.
 To see
 this, consider the interface between a void and the surrounding sea
 of particles with at least $k$ overlapping neighbors.  For a
 sufficiently large void, a particle on the boundary with at least $k$
 overlapping neighbors will inevitably experience a nonzero net force
into the void, since floaters provide no
 compensating force.  Thus, large voids will shrink away;
 they will not grow.  To capture some of this physics, we have introduced
 a 3-core model with `''force-balance'' on the square
 lattice~\cite{schwarz.liu.chayes}.  Potential neighbors
 are located within a $5$x$5$ square centered upon the site of
 interest;
 thus, $Z_{max}=24$.  To survive decimation, any given site must have at 
least $3$ occupied neighbors and if there is at least one
 occupied neighbor to the
 right of the site of interest, there must be at least one occupied
 neighbor to its left and vice versa. Similarly, if there is at least one
occupied
 neighbor above the site of interest, there must be at least one occupied 
 neighbor below and vice versa.

 We have undertaken simulations of this $3$-core model, and find
 evidence that the transition is discontinuous with a
 diverging correlation length.  In Fig. 1a, the fraction of
 occupied sites in the spanning cluster, $\kappa$, is
 plotted
 as a function of $p$ for different
 system lengths $L$.  For each $L$ we observe
 that $\kappa$ jumps from zero to $\kappa_c(L)$ at some $p_c(L)$.
 For a continuous transition, the jump $\kappa_c(L)$ would decrease with
 $L$ and vanish as $L \rightarrow \infty$, but here
 $\kappa_c(L)$ {\it increases} with $L$.
The size of the jump $\kappa_c(L)$ appears to approach a nonzero
limiting value of $0.374(1)$  (solid triangles in inset to Fig. 1b).  
Furthermore, 
the transition point, $p_c(L)$ approaches $p_c=0.396(1)$ for $L \rightarrow \infty$.
To verify that unstable voids do not drive the transition to $p_c=1$, we plot $1-p_c(L)$
vs. $1/\ln L$ as open diamonds in the inset to Fig. 1b.  We do not find linear behavior with a
$y$-intercept of zero, as predicted for $p_c=1$~\cite{aizenmann.lebowitz}.

 We calculate the correlation length exponent
 from two different quantities (Fig. 1b): (1) the width of the
 transition defined by $W=p_{+}(L)-p_{-}(L)$, where $p_{\pm}(L)$ are
 defined as the values of $p$ at which
 the probabilities of obtaining a spanning cluster are $0.25$
 and $0.75$, respectively, and (2) the critical point shift $\delta p_c \equiv p_c-p_c(L)$.  
Both $W$ and $\delta p_c$ scale as $L^{-1/\nu}$ with
 $\nu=1.52(2)$ and $1.53(1)$ respectively, as shown in Fig. 1b.  This exponent
leads to scaling collapse of the order parameter curves of Fig. 1a, as
shown in the inset to Fig. 1a.  Here, we assume the scaling form $\kappa(p,L)=
 \kappa_c(L)+L^{-\beta/\nu}f((p-p_c(L))L^{1/\nu})$, with
 $\beta=1.0$ (for optimal collapse).
 For ordinary first-order transitions, finite-size
scaling would predict a diverging length with an exponent of $1/d$~
\cite{fisher.berker}, corresponding to $0.5$ in $d=2$.  We obtain
an very different exponent, strongly suggesting a mixed transition.  Analysis of a recently-proposed lattice 
model reaches a similar conclusion~\cite{recent}.  Furthermore, a recent $1/d$
 expansion of pure $k$-core percolation suggests that the
 mixed nature of the mean field transition may survive in finite 
 dimensions~\cite{harris.schwarz}.

 While $k$-core percolation appears to capture the minimal physics
needed to
 explain the mixed transition found at Point J, it is not a complete
 description of jamming.  This can already be seen by comparing the
 exponents observed for the mixed transition of the $d=2$ $3$-core
 model, $\beta \approx 1.0$ and $\nu \approx 1.5$, with those found in the $d=2$ and
 $d=3$ jamming simulations, $\beta=0.49$ and $\nu=0.24$.  In fact, $k$-core models do not include a very important property of the jamming
 transition, namely isostaticity\cite{shlomo}.
 At Point J, the number of overlapping
 neighbors jumps
 from zero to $Z_c=2d$ where $d$ is the dimensionality.  
 In $k$-core
 percolation, on the other hand, $Z_{c}$ is not universal; it depends on
 $k$ and $Z_{\max}$.  We find that the global constraint of
 $k$-core percolation yields $Z_{c}>k$, even though the local
 constraint only requires $k$ neighboring occupied sites per site.

Duxbury, {\it et al.}\cite{rigidity} have proposed that the closely-related
 problem of rigidity percolation can be mapped onto the $k$-core
 percolation problem in mean-field by imposing the constraint that the
 transition should occur when $Z_c$ satisfies the isostatic
 condition.  Thus in their formulation, the mean-field transition occurs
above $p_c$, at some $p_r$ at which $Z_c$
 reaches its isostatic value.  They therefore obtain an isostatic, ordinary first-order
transition, while $k$-core percolation yields a non-isostatic, mixed
transition.  Neither case properly applies to the
 jamming transition, which appears to be both mixed and
isostatic~\cite{ohern.silbert.liu.nagel,silbert.liu.nagel}.
 We note that a complementary theory by Wyart, {\it et
 al}.~\cite{wyart.nagel.witten} starts with isostaticity at Point J
 and successfully describes the
 behavior of the
 density of vibrational modes and predicts a diverging length scale 
with exponent of 1/2, with much the same physical meaning as our
 $\nu^{\#}=1/2$.  In addition, a recent field theoretical approach also
   starts with isostaticity and appears to produce some of the same
 mean-field
 exponents that we find~\cite{bulbul}.  We speculate that a complete
 theory of jamming would exhibit the same mean-field exponents as $k$-
 core percolation, but different behavior in finite
 dimensions due to isostaticity; the latter effect suggests an upper
 critical dimension of two~\cite{wyart.nagel.witten,wyart.silbert.nagel.witten}.

 We have argued that the physical constraint of requiring at least
 $k=d+1$
 overlapping neighbors per particle in zero-temperature sphere packings
 leads to a transition resembling the $k$-core percolation type.
 However,
 this analogy may have implications ranging beyond sphere packings
 to glassforming liquids.  This connection is suggested by the
 set of exponents
 we find for mean-field $k$-core percolation, which is rare but has been seen 
in a few
 other models that are known to exhibit glassy dynamics as the
 temperature
 is lowered.  These include
 the mode-coupling theory of
 glasses~\cite{biroli.bouchaud}, mean-field theories of
 the $p$-component spin
 glass~\cite{kirkpatrick.wolynes,kirkpatrick.wolynes2}  and
 kinetically-constrained spin
 models~\cite{fredrickson.andersen,kob.andersen,toninelli.biroli.fisher,
 sellitto.biroli.toninelli}.
 For the latter models, this is no coincidence since they map onto
 $k$-core percolation and its 
 variants~\cite
 {kob.andersen,toninelli.biroli.fisher,sellitto.biroli.toninelli}.
Finally, we note that models such as the 3-SAT spin glass are also
variants of $k$-core percolation.  It can be shown that in mean field, the
unfrustration-frustration transition has the same mixed character as
in $k$-core percolation, suggesting that the 3-SAT spin glass
may exhibit glassy dynamics.

 It has been proposed~\cite{liu.nagel,liu.nagel.book} that the behavior
 of many jamming
 systems, including glasses, suspensions, foams and granular materials,
 might be captured by ``jamming phase diagrams,'' in the
 three-dimensional space of temperature $T$,  applied
 shear stress $\sigma$, and packing density $\phi$. In this space, the
 boundary separating jammed from unjammed behavior
 is a ``surface" whose location is nebulous because it depends on the
 time-scale of the observations, and Point J lies underneath the jamming
 surface.
 Numerical simulation results~\cite{ohern.silbert.liu.nagel,silbert.liu.nagel}
 suggest
 that the entire jamming surface of the jamming phase diagram is indeed
 controlled by
 Point J, the unique point
 where a sharp transition occurs.  Here, we have argued that the physics
 near Point J is
 strongly suggestive
 of the $k$-core problem.  The latter has a
 transition with unusual features
 that
 mirror corresponding features found at Point J:  a mixed first--second
 order transition and
 the same exponents that characterize
 the continuous part.  In contrast to earlier scenarios of the glass
 transition
 based on avoided
 critical points, either at nonzero temperature~\cite{kivelson.tarjus}
 or zero temperature~\cite{sethnaandco,berthierandco}, the arguments of
 O'Hern, {\it et al.}
 \cite{ohern.silbert.liu.nagel} evidently suggest a
 scenario based on an avoided {\it mixed} transition at Point J.
 On
 one hand, the first-order
 character of the transition may explain the presence of strong
 system-specific features such as
 the degree of fragility.
 On the other hand,
 the continuous component of the transition may
 explain the many
 ubiquitous features in the phenomenology of
 jamming~\cite{liu.nagel.book}.

 %\acknowledgements

 We thank A. B. Harris, T. C. Lubensky, S. R. Nagel and M. Wyart for instructive
 discussions, and also G. Biroli and D. S. Fisher for communication and cooperation.  
We are grateful for the support of
 NSF-DMR-0087349 (AJL,JMS), DE-FG02-03ER46087 (AJL,JMS) and
 NSF-DMS-0306167 (LC).

\end{document}